\documentclass{IOS-Book-Article}
\usepackage{fancyvrb}

\usepackage{mathptmx}
\usepackage{soul}\setuldepth{article}

\usepackage{hyperref}
\usepackage{color}
\usepackage{mdframed}
\usepackage{graphicx}
\usepackage{float}
\usepackage{multirow}
\usepackage{tabularx}
\usepackage[table]{xcolor}
\usepackage{colortbl}
\usepackage{amsmath}
\usepackage{rotating} 
\usepackage{makecell} 
\usepackage{url}
\usepackage{graphicx}
\usepackage{array}
\usepackage{colortbl}
\usepackage{booktabs}
\usepackage{array}
\usepackage{multirow}
\usepackage{xcolor}
\usepackage{soul}
\sethlcolor{blue!15}
\newcommand{\hlorange}[1]{{\sethlcolor{orange!15}\hl{#1}}}

\usepackage{listings}

\usepackage[most]{tcolorbox}
\usepackage[table]{xcolor}
\usepackage{tikz}

\definecolor{UserColor}{rgb}{0.6, 0.6, 0.6}
\definecolor{GPTColor}{rgb}{0.8, 0.8, 1}

\newtcolorbox{myblock}[1]{colback=UserColor!10, colframe=UserColor, coltitle=white, title=#1, fonttitle=\bfseries, breakable}

\definecolor{responseBlock}{RGB}{232,217,245}
\definecolor{responseBlockFrame}{RGB}{135,80,184}
\newtcolorbox{myexampleblock}[1]{
colback=GPTColor!10, 
colframe=GPTColor, 
coltitle=white, 
title=#1, 
fonttitle=\bfseries,
breakable}

\newtcolorbox{exampleblock}[1]{colback=teal!25, colframe=teal, coltitle=white, title=#1, fonttitle=\bfseries}

\def\hb{\hbox to 11.5 cm{}}

\begin{document}

\pagestyle{headings}
\def\thepage{}
\begin{frontmatter}

\title{Robots in the Middle: \\
Evaluating LLMs in Dispute Resolution}

\markboth{}{September 2024\hb}

\author[A]{\fnms{Jinzhe} \snm{Tan}
\thanks{Corresponding Author: Jinzhe Tan, University of Montreal, jinzhe.tan@umontreal.ca.}},
\author[B]{\fnms{Hannes} \snm{Westermann}},
\author[C]{\fnms{Nikhil Reddy} \snm{Pottanigari}},
\author[D]{\fnms{Jaromír} \snm{Šavelka}},
\author[A,E]{\fnms{Sébastien} \snm{Meeùs}},
\author[A]{\fnms{Mia} \snm{Godet}},
and
\author[A]{\fnms{Karim} \snm{Benyekhlef}},

\runningauthor{J. Tan et al.}
\address[A]{Cyberjustice Laboratory, University of Montreal, Canada}
\address[B]{Maastricht Law and Tech Lab, Maastricht University, Netherlands}
\address[C]{Mila - Quebec AI Institute, University of Montreal, Canada}
\address[D]{School of Computer Science, Carnegie Mellon University, United States}
\address[E]{Faculty of Law and Criminology, Université Libre de Bruxelles, Belgium}

\begin{abstract}
Mediation is a dispute resolution method featuring a neutral third-party (mediator) who intervenes to help the individuals resolve their dispute. In this paper, we investigate to which extent large language models (LLMs) are able to act as mediators. We investigate whether LLMs are able to analyze dispute conversations, select suitable intervention types, and generate appropriate intervention messages. Using a novel, manually created dataset of 50 dispute scenarios, we conduct a blind evaluation comparing LLMs with human annotators across several key metrics. Overall, the LLMs showed strong performance, even outperforming our human annotators across dimensions. Specifically, in 62\% of the cases, the LLMs chose intervention types that were rated as better than or equivalent to those chosen by humans. Moreover, in 84\% of the cases, the intervention messages generated by the LLMs were rated as better than or equal to the intervention messages written by humans. LLMs likewise performed favourably on metrics such as impartiality, understanding and contextualization. Our results demonstrate the potential of integrating AI in online dispute resolution (ODR) platforms. 

\end{abstract}

\begin{keyword}

large language models\sep artificial intelligence\sep online dispute resolution\sep access to justice\sep ai \& law\sep chatgpt

\end{keyword}
\end{frontmatter}
\markboth{September 2024\hb}{September 2024\hb}

\section{Introduction}

Intermediaries (for example, mediators, arbitrators, or conciliators) can play an important role in facilitating dispute resolution. When a discussion turns emotionally charged, communication breaks down, or the dispute reaches a deadlock, intermediaries can intervene with information to help calm emotions, clarify facts, identify the key issues in the dispute, and make proposals for settlement, thereby promoting the satisfactory progress of dispute resolution.

Of course, the involvement of such intermediaries is limited to areas where the value for the dispute is higher than the cost of the intermediary. Further, in some areas, there may simply not be a sufficient number of trained facilitators to cover all of the disputes \cite{branting2023computational}. Supporting mediation through technological tools is thus a promising avenue of increasing the scalability of facilitated dispute resolution, and enabling its use in new contexts.

The recent advancements in large language models (LLMs) have opened the door to the use of AI to assist intermediaries in understanding dispute scenarios, offering AI-suggested interventions, and even AI automated interventions \cite{westermann2023llmediator}. However, the complex, interactive and interpersonal nature of dispute resolution sets a high bar for such tasks \cite{larson2010artificial}. Intermediaries need a nuanced skill set, including contextual understanding, emotion perception, and the ability to propose balanced, contextually appropriate solutions. While LLMs have demonstrated considerable capabilities in discrete tasks (such as contextual awareness and language understanding), their performance in more complex, integrated tasks remains under-explored.

To investigate the performance of large language models (LLMs) like \texttt{GPT-4o} in dispute resolution tasks, we analyzed their abilities in selecting \textit{intervention types} and generating \textit{intervention messages} based on \textit{dispute scenarios} from a novel corpus of 50 hypothetical disputes. Based on these disputes, we investigate three research questions:

\begin{enumerate}
  \item[RQ1] To what extent can LLMs select appropriate intervention types given a dispute scenario?
  \item[RQ2] How do LLMs compare to humans in drafting intervention messages?
  \item[RQ3] To what degree are messages generated by LLMs safe and free of hallucinations?
\end{enumerate}

\section{Related Work}

The use of computational methods to facilitate dispute resolution is a long-standing topic in the field of Legal Informatics, such as in the ICANS system, where the parties can choose their preferences through a mathematical mechanism and gradually reach an agreement with the assistance of the system \cite{thiessen1993icans}. Using a similar idea, Family\_winner uses a game-theoretic approach that allows users to split up and resolve disputes using repeated offers \cite{bellucci2001representations,zeleznikow2003family_winner,bellucci2005developing}. Other approaches include indicating potential court outcome ranges to align the expectations of the parties \cite{carneiro2014online,zeleznikow2016can,susskind2019online,benyekhlef2018intelligence}). These approaches have laid the groundwork for applying technology to the mediation process.

In recent years, with advances in natural language processing (NLP), discussions and attempts to use language models to facilitate dispute resolution have emerged
\cite{larson2010artificial, larson2010brother}. For example, Branting et al. use the example of Utah's ODR system to analyse how language models can be used to analyse the stages of a dispute and provide facilitators with recommendations based on standard text message \cite{branting2023computational}. 

The evolution of LLM marks a significant shift from earlier domain-specific models. LLMs have achieved significant performance in terms of their foundational capabilities \cite{achiam2023gpt}, contrary to conventional models that are specific to particular domains. LLMs excel in tasks such as, e.g., language understanding and generation \cite{chen2023robust}, sentiment analysis \cite{zhang2023sentiment}, and reasoning \cite{huang2022towards}. These foundational capabilities allow LLMs to be adapted to various domains through techniques such as fine-tuning or prompt engineering. This flexibility has already led to diverse applications in the legal field, including providing legal information \cite{tan2023chatgpt,westermann2023bridging}, acting in fiduciary roles \cite{nay2023large}, conducting empirical research \cite{drapal2023using}, analyzing legal text data \cite{savelka2023can,savelka2023explaining}, and developing legal expert systems \cite{janatian2023text}.

The main direction of our exploration in this paper is how well LLMs such as \texttt{GPT-4o} perform in selecting intervention types and generating intervention messages, rather than expecting to deploy them directly in practical applications.

\section{Proposed Framework}
\label{sec:prop_framework}

We use the LLMediator framework presented in \cite{westermann2023llmediator} to set up a dispute scenario involving two disputing parties and a mediator. The parties can communicate with each other through text messages, and the mediator can intervene in the dispute through intervention messages in order to assist the disputants in reaching a resolution. In the LLMediator framework, this functionality can be performed by a human, by a human assisted by an LLM, and potentially in a fully automated fashion \cite{westermann2023llmediator}. In this paper, we investigate how human-written messages compare to those generated by LLMs.

\begin{figure}[ht]
    \centering
    \includegraphics[width=0.6\linewidth]{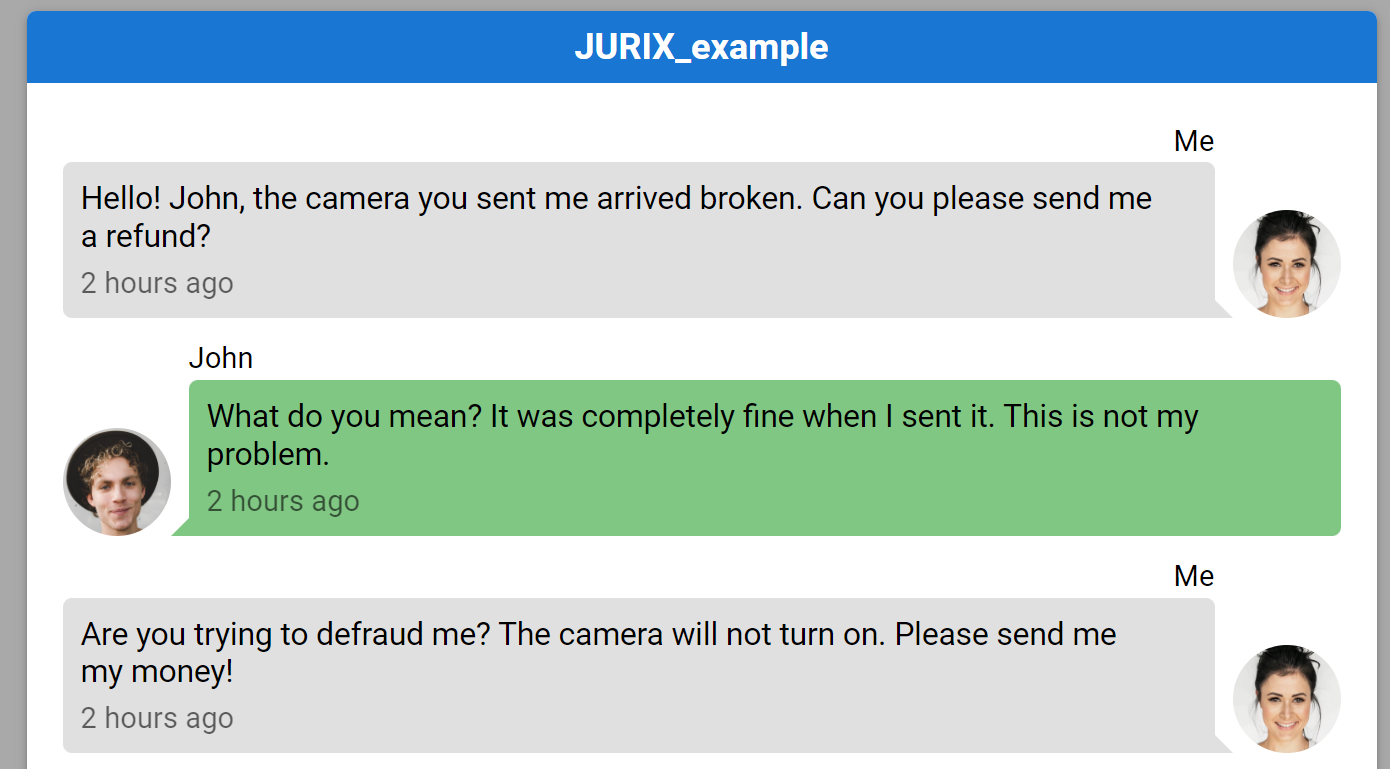}
    \caption{A screenshot from the LLMediator, showing a dispute prior to the mediator's intervention.}
    \label{fig:interface}
\end{figure}

Figure \ref{fig:interface} shows a screenshot of a dispute. At this point, the mediator may decide that it is time to intervene, in order to help the disputants find an amicable solution. In intervening, the mediator needs to perform two important steps:

\textbf{Step 1 - Decide intervention type.} Depending on the state of the discussion, the mediator may want to include different types of interventions in the messages they send. For example, they may want to calm down the discussion, encourage exchanges of information or help the parties evaluate their alternatives. We adopted a list of types of possible interventions from \cite{Dispute_Resolution_Reference_Guide_2007}, as can be seen in table \ref{tab:actions}. Deciding on the type of intervention requires an understanding of the dispute context and empathy towards the parties.

\textbf{Step 2 - Draft the intervention message.} The mediator has to decide which specific words to use to perform the type of intervention they have chosen. Clearly and empathically communicating these ideas is important to help the parties achieve their goals, while avoiding mistakes and ambiguities.

These steps need to be carried out implicitly by the mediator intervening in a dispute. We chose to adopt them as a framework to compare the messages created by human annotators and LLMs. By splitting the process into these two steps, we can compare the performance of the two across multiple steps, giving us a deeper understanding of the difference.

\section{Experimental Design}

Communications between parties during disputes often involve sensitive and private information leading to a scarcity of accessible data on dispute scenarios. 
Therefore, we manually constructed a dataset comprising 50 \textit{dispute scenarios} for our experiments (section \ref{sec:dispute_construction}). Afterwards, human mediators and LLMs intervened in the dispute scenarios (sections \ref{subsec:Intervention_by_human} \& \ref{sec:llm_interventions}), following the same instructions, and both were assessed in a blind evaluation for their performance in the same scenarios (section \ref{sec:evaluation}).

\subsection{Constructing disputes}
\label{sec:dispute_construction}

The 50 dispute scenarios we created followed the same structure, with each scenario consisting of two dialogues between Party A and Party B, thus featuring a total of four textual messages. In order to ensure a diverse set of disputes, we wrote disputes with varying characteristics, as described in Table \ref{tab:disputes}.

\begin{table}[ht]
\centering
\begin{tabular}{|p{1.3cm}|p{3.8cm}|p{5.5cm}|}
\hline
\textbf{Type} & \textbf{Explanation} & \textbf{Example} \\
\hline
\textbf{Emotional} & The parties have strong emotional expressions in the conversation. & A person asks their neighbour to keep their dogs quiet, resulting in an escalating conversation with threats. \\
\hline
\textbf{Complex} & The dispute has a high degree of complexity and the facts of what happened are difficult to clarify. & A person asks an insurance company to pay for a car accident, resulting in a discussion of legal and technical nuances. \\
\hline
\textbf{Confusion} & The parties are confused, leading to difficulties in communication.  & A  customer and merchant disagree on the details of an undelivered order, leading to repeated requests for more information. \\
\hline
\textbf{Impossible} & The dispute features strong disagreements, resulting in a deadlock. & A customer requests a laptop to be repaired, but the manufacturer argues that the damage is cause by the user, refusing the warranty.  \\
\hline
\textbf{Evidential} & The dispute centers around conflicting evidence or claims. & One party insists that an agreement regarding a computer sale was reached, while the other disagrees. \\
\hline
\end{tabular}
\caption{Description of dispute characteristics}
\label{tab:disputes}
\end{table}

After reviewing the dispute scenarios, we found that they were diverse both in terms of communication style and legal areas, covering areas such as parcel delivery, land property rights disputes, noise complaints, and so on. This diversity contributes to helping us perform robust evaluations of the interventions.

\subsection{Human interventions}\label{subsec:Intervention_by_human}

After creating the dispute scenarios, we manually created interventions for the disputes. We randomly assigned our annotators (all of whom are co-authors on this paper, with varying legal knowledge) to the disputes. For each dispute, we asked the annotators to perform the steps described above in section \ref{sec:prop_framework}. First, we thus asked them to select from one to three of the intervention types described in table \ref{tab:actions}. Second, we asked the annotators to draft intervention messages based on the chosen intervention types. For coherence with the LLM written messages, we asked the mediators to maintain a one-to-one correspondence between the chosen intervention types and the generated message, by writing 1-2 sentences for each intervention type to construct their message. Table \ref{tab:example} shows an example of chosen intervention types and resulting messages by humans and LLMs.

\begin{table}[htbp]
\centering
\begin{tabular}{|p{2.2cm}|p{3.8cm}|p{5cm}|}
\hline
\textbf{Dispute summary} & \textbf{Intervention types chosen} & \textbf{Interventions} \\
\hline
Party A asks Party B to delete their picture from social media. B refuses and makes fun of A. & 
\textbf{Human}: \hl{4. Promote a productive level of emotional expression}, \hlorange{13. Propose solutions that meet the fundamental interests of all parties.} \textbf{(preferred)} & 
\textbf{Human:} \hl{Let's not insult each other or downplay anyone's feelings.} \hlorange{B, you must delete the picture, since keeping it without A's consent is illegal.} \\
\cline{2-3}
 & \textbf{LLM:} 2. Help the parties understand each other's views, 
3. Let the parties know that their concerns are understood, 9. Encourage flexibility and creativity & 
\textbf{LLM:} \hl{Let's ensure the conversation is respectful.} \hlorange{Here's a proposition: Party B, how about posting an image that both of you find humorous and enjoyable instead? This way, Party A won't feel embarrassed and both of you can have fun.} \textbf{(preferred)} \\
\hline
\end{tabular}
\caption{Example of selected intervention types and written interventions, both by human and LLM. Here, the evaluator preferred the human choice of intervention types, but the LLM-generated intervention message.}
\label{tab:example}
\end{table}

\subsection{LLM Interventions}
\label{sec:llm_interventions}

For LLMs, we followed the same process as for human interventions, of first using the LLM to choose an intervention type, and then generating an intervention message based on chosen intervention types. We used the \texttt{gpt-4o-2024-05-13} model via the \texttt{openai} python Python library\footnote{Github: OpenAI Python Library. Available at: \url{https://github.com/openai/openai-python} [Accessed 2024-08-26]}, which was the state-of-the-art model at the time of the experiment. During the experiment, we used the default parameters.

\textbf{Step 1 - Decide intervention type.} \label{sec:Generate_intervention_type} First, we asked the model to select between one and three intervention types to respond to a provided dispute. We used the mediator's guide given on the website of Department of Justice of Canada \cite{Dispute_Resolution_Reference_Guide_2007} to create the prompt, which covers the disputed conversation and the 13 types of interventions (see table \ref{tab:actions}).

\textbf{Step 2 - Generate intervention message.} \label{sec: Generate_intervention_message} We then provided \textit{human-selected} intervention types as inputs to the models and asked them to write intervention messages based on the intervention types. The LLM was also asked to maintain the correspondence between the intervention types and the text (see section \ref{subsec:Intervention_by_human} and table \ref{tab:example}). 

We always use the intervention types chosen by the human annotator in order to be able to compare the quality of the written intervention. Thus, we are able to compare LLMs to humans on two tasks: choosing the correct intervention types, and generating an intervention message based on chosen intervention types. Our choice of using the human messages does not imply that we considered the intervention types selected by the humans to be superior to those selected by the LLMs---in fact, in the evaluation, we found that the evaluators often preferred the intervention types selected by the LLMs.

\begin{table}[ht]
\centering
\begin{tabular}{|c|p{11cm}|}
\hline
\textbf{No.} & \textbf{Intervention Types} \\
\hline
1 & Encourage exchanges of information \\
\hline
2 & Help the parties understand each other's views \\
\hline
3 & Let the parties know that their concerns are understood \\
\hline
4 & Promote a productive level of emotional expression \\
\hline
5 & Lay out the differences in perceptions and interests \\
\hline
6 & Identify and narrow issues \\
\hline
7 & Help parties realistically evaluate alternatives to settlement \\
\hline
8 & Suggest that the parties take breaks when negotiations reach an impasse \\
\hline
9 & Encourage flexibility and creativity \\
\hline
10 & Shift the focus from past to future \\
\hline
11 & Shift the focus from one of blame to a creative exchange between the parties \\
\hline
12 & Hold caucuses with each disputant if there is deadlock or a problem \\
\hline
13 & Propose solutions that meet the fundamental interests of all parties \\
\hline
\end{tabular}
\caption{List of intervention types to facilitate mediation from \cite{Dispute_Resolution_Reference_Guide_2007}}
\label{tab:actions}
\end{table}

\subsection{Evaluation}
\label{sec:evaluation}

\textbf {E1 - Evaluation of intervention types.} After obtaining the types of intervention chosen by humans and LLMs based on the dispute scenarios according to the process described in Section \ref{sec:Generate_intervention_type}, we conducted a blind evaluation on the choice of type. We asked evaluators to compare the two intervention type choices after reading the dispute scenario and to judge which choice they thought was superior on a  5 point Likert scale. 

Although there are multiple reasonable intervention type choices for each dispute scenario, some options may be more suitable depending on the context. For example, if the parties use impolite language or express strong emotions, selecting intervention type No. 4, `Promote a productive level of emotional expression,' would be more appropriate. In situations where negotiations are deadlocked, choosing intervention type No. 8, `Suggest that the parties take breaks when negotiations reach an impasse,' or No. 12, `Hold caucuses with each disputant if there is deadlock or a problem,' would be more fitting.

\textbf{E2 - Evaluation of intervention messages.} Afterward, we assigned another evaluator to each dispute. We asked them to assess (in a blind fashion) which of the two intervention messages they preferred. The evaluators first provided their overall preference using a 5 point Likert scale, and wrote comments motivating their choice. Then, they compared the messages in terms of specific rubric items, including understanding and contextualization, neutrality and impartiality, empathy awareness, and resolution quality. After completing the evaluation based on these criteria, the evaluators were asked to write additional notes highlighting any noteworthy points.

\textbf{E3 - Safety evaluation of LLM interventions.} Finally, we conducted safety and quality checks of the messages generated by the LLMs. We assessed whether the model hallucinated and whether there were any safety issues with the generated messages.

\section{Results}

\subsection{E1 - Evaluation of intervention types}

\begin{figure}[h!]
    \centering
    \includegraphics[width=0.8\textwidth]{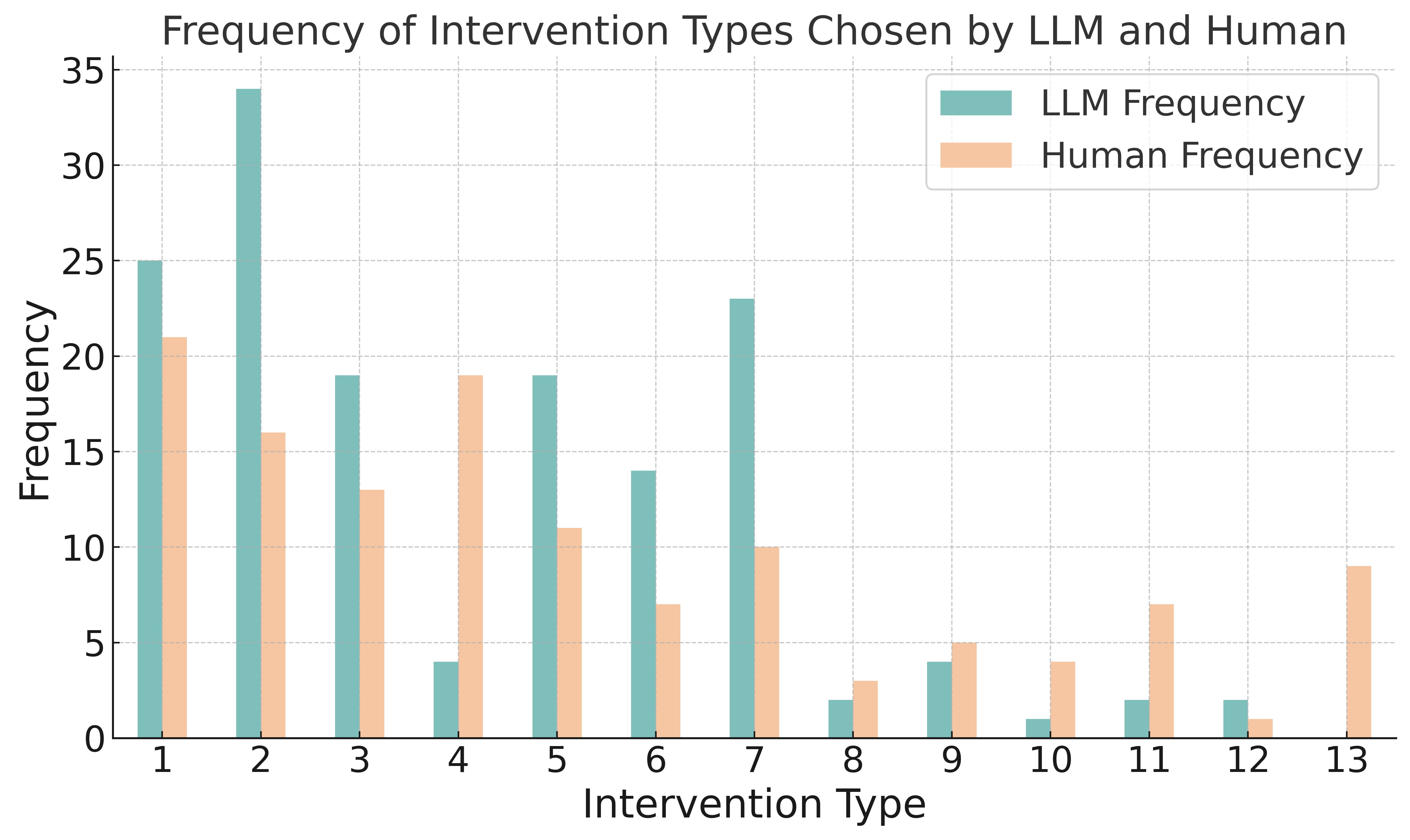}
    \caption{Frequency of Intervention Types Chosen by LLM and Human}
    \label{fig:intervention_types_chart}
\end{figure}

Table \ref{table:comparison_types} shows the results of the blind evaluation. We found that evaluators generally preferred the intervention types chosen by LLMs. However, there were instances when these choices showed strong variance. Figure \ref{fig:intervention_types_chart} shows the distribution of interventions chosen by humans, compared to those suggested by LLMs.

\definecolor{custom1}{HTML}{7FBFBB}
\definecolor{custom2}{HTML}{96BCBD}
\definecolor{custom3}{HTML}{C7E1D8}
\definecolor{custom4}{HTML}{F6C5A1}
\definecolor{custom5}{HTML}{F2A778}

\newcommand{\circlemark}[1]{\raisebox{-.5ex}{\tikz\draw[#1,fill=#1] (0,0) circle (0.15cm);}}

\begin{table}[ht]
\centering
\begin{tabular}{|l|c|}
\hline
\textbf{Description} & \textbf{Number of responses} \\ 
\hline
\circlemark{custom1} LLM is significantly better than Human & 11 \\ 
\hline
\circlemark{custom2} LLM is slightly better than Human & 11 \\ 
\hline
\circlemark{custom3} LLM and human are about the same & 9  \\ 
\hline
\circlemark{custom4} Human is slightly better than LLM & 14 \\ 
\hline
\circlemark{custom5} Human is significantly better than LLM & 5  \\ 
\hline
\end{tabular}
\vspace{8pt} 
\caption{We used a 5 point Likert scale to compare human evaluators' preferences for LLM and human-selected intervention types.}
\label{table:comparison_types}
\end{table}

\subsection{E2 - Evaluation of intervention messages}

\begin{figure}[ht]
    \centering
    \includegraphics[width=\textwidth]{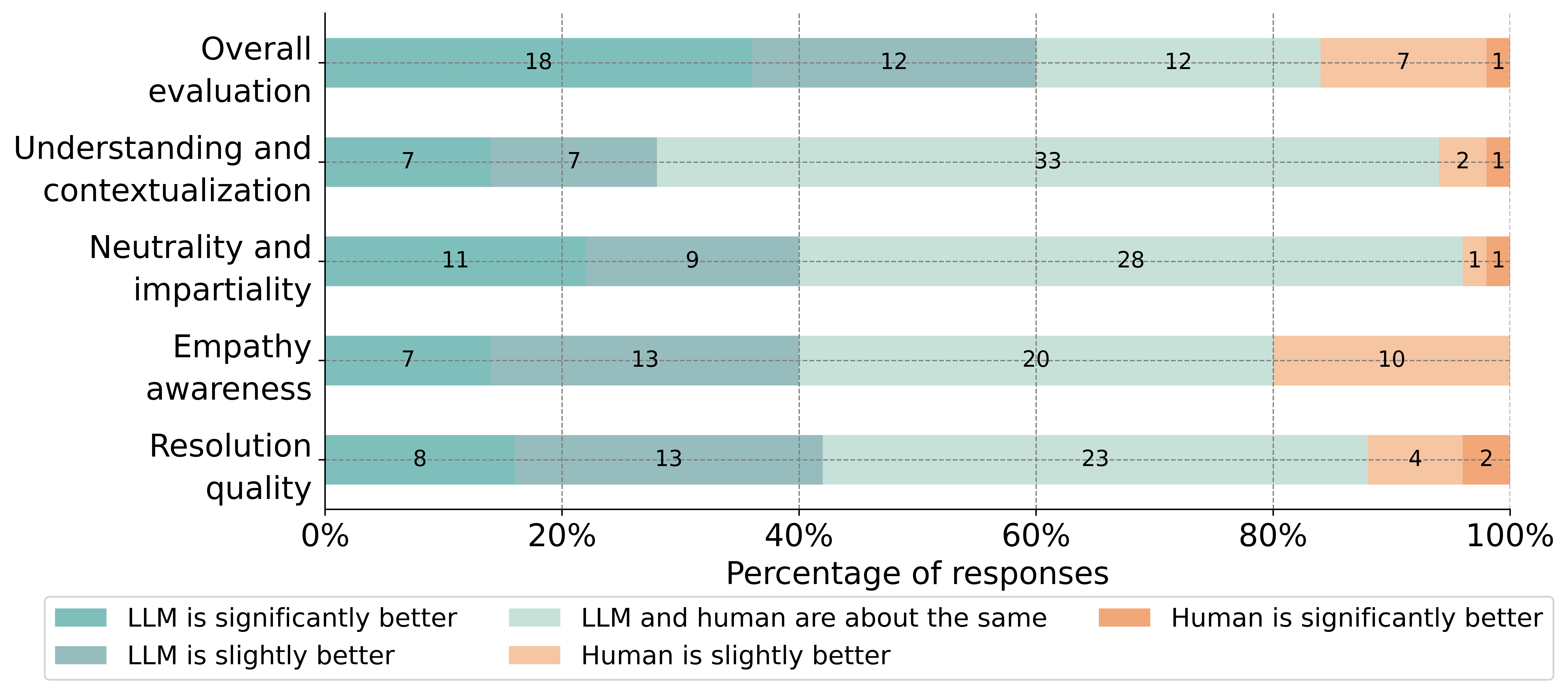} 
    \caption{The bar chart shows the distribution of responses evaluating the performance of LLMs compared to humans across the five metrics we set.}
    \label{fig:messages}
\end{figure}

Figure \ref{fig:messages} shows the blind evaluation preference of the evaluators on the different axes between the human and LLM-generated messages. As we can see, there was a strong preference for the messages written by the LLMs, across the different categories. 

In terms of overall evaluation, 84\% of evaluators believed that the intervention messages generated by LLMs were either superior or equivalent to those created by human mediators, with LLMs significantly or slightly outperforming humans in 60\% of cases. Further, the LLM-generated messages were scored as equal to or better than the human messages in between 80\% and 96\% of the cases in all categories (see figure \ref{fig:messages}).

\subsection{E3 - Safety evaluation of LLM-generated intervention}

After manually checking all the LLM-generated messages, we did not find the appearance of harmful messages and hallucinations in the scenario of this experiment. However, this result does not guarantee that the LLM will always be safe in larger-scale experiments or real-world applications, it simply indicates that no such phenomena were detected in this particular experimental scenario.

\section{Discussion}

\subsection{RQ1 - To what extent can LLMs comprehend dispute scenarios and select appropriate intervention types?}
Table \ref{table:comparison_types} shows that our human evaluators preferred the intervention types chosen by the LLMs in 22 cases, were ambivalent in 9 cases, and preferred the human-chosen types in 19 cases. Overall, this is a strong result suggesting viability of LLMs on this complex task, requiring the nuanced understanding of a dispute and empathy to determine which steps to take next. At the same time, it should be noted that the evaluators were not expert mediators, and that the task of determining which specific intervention type is appropriate may be subjective (see section \ref{sec:limitations}).

Figure \ref{fig:intervention_types_chart} shows the distribution of intervention types chosen by humans and LLMs. Here, differing patterns are revealed. The top three types of interventions chosen by the LLMs are helping the parties to understand each other's views, the encouragement of exchanging information and the helping of parties to evaluate alternatives (2,1,7). The human annotators, on the other hand, preferred the encouragement of exchanging information, the promotion of a productive level of expression, and helping the parties understand each others views (1,4,2).

These preferences may reveal a different understanding of what is important in mediation. At the same time, prior work has shown that LLMs may be affected by the order of presented options \cite{pezeshkpour2023largelanguagemodelssensitivity}. This may partially explain why the LLMs seem to prefer the early items in the list, although a similar preference also seems present for the human mediators. 

\subsection{RQ2 - How do LLMs compare to humans in drafting intervention messages?}

Our experiment shows that LLMs can perform at a level comparable to or even better than our human annotators. The LLM-generated messages were rated higher or equal to the human-written ones in 84\% of the scenarios. While certain caveats apply (see section \ref{sec:limitations}), these results highlight the impressive capability of LLMs in drafting appropriate intervention messages. Various reasons were given as to why the evaluators preferred the messages written by the LLM. 

\emph{First}, the evaluators often found the messages written by the LLM to be more smooth and clear than the human-written ones. The general tone used by LLMs, involving frequent messages such as ``I completely understand'' or ``It seems like there are problems,'' seems to work well in a mediation environment, and may have contributed to high scores.

\emph{Second}, while LLMs are known to frequently ``hallucinate''  information \cite{das2023diving, dahl2024large}, in our case the humans more often misunderstood the dispute or were confused about the party intentions or factual occurrences. This could be due to factors such as fatigue, emotional bias, or a misunderstanding of the role of the mediator. In contrast, LLMs demonstrated consistent and coherent interventions across multiple cases, with fewer instances of judgment errors. 

\emph{Third}, we found that our human annotators would often propose very specific solutions or even indicate fault, which received a lower rating as it may not be appropriate for the role of the mediator. 

Overall, while it is important to highlight the caveat of none of the annotators and evaluators having experience in mediation, it seems like the messages generated by the LLMs capture the dispute well, use an appropriate tone, are clear and do not overreach, making them compare favourably to the messages written by our human annotators.

\subsection{RQ3 - To what degree are messages generated by LLMs safe and free of hallucinations?}

We did not notice any unsafe messages or hallucinated information in the generated messages. While this of course does not rule out such issues, it is nonetheless a promising result for the use of LLMs in a dispute resolution context. The approach discussed in \cite{westermann2023llmediator}, where the generated messages are provided to a human mediator before being sent to the parties, could further mitigate such concerns.

\subsection{On the use of gold-standard data}
\label{sec:gold}
Using human-generated answers as Ground Truth (`gold standard') is a very common practice in machine learning research, which helps us create benchmarks for evaluating the performance of algorithms or models. Here, we took a different approach, instead asking to compare human-generated messages to LLM-generated ones, thereby not assuming that the human data can serve as a reliable gold standard. With good reason - looking at the results, the LLM generated messages were consistently rated higher than the messages written by the humans.

However, the results also reveal the general difficulty of evaluating the performance of models that can perform complex, nuanced tasks without giving obviously wrong answers. None of the messages written by the LLM contained any hallucinations or other obvious defects, which makes the overall assessment difficult and subjective. Perhaps, as discussed in \cite{ma2023conceptual}, it is more useful to see the annotations as surveys of individual views, rather than a single ``truth'', when it comes to bespoke and nuanced legal tasks. Regardless, the science of evaluating large language models on legal tasks is in its infancy, and we hope that this paper can contribute some insights to this important issue.

\section{Limitations}
\label{sec:limitations}

In this work, we used a structured evaluation method to compare the performance of LLMs to humans. While the results are promising, there are some important caveats. \emph{First}, the process of selecting an intervention type and then being bound to it may not correspond to how mediators approach drafting messages in reality. Likewise, the drafting of messages in blocks organized by intervention types may also impose artificial limitations on the types of interventions that can be written. 

\emph{Second}, our disputes and messages were drafted and evaluated by people without specific training in mediation, and none of whom are native English speakers. This may give an advantage to the LLMs. While it seems like the ability of the LLM to select intervention types and write messages is favourable to that of average people, this paper cannot tell us about how trained mediators would approach these issues.

\emph{Third}, as touched upon in section \ref{sec:gold}, it may not be possible to assess which intervention type or message is ``better'' without observing real-world outcomes, leading to a subjective assessment. For example, it is possible that grammar mistakes and our expectations of the tone of the mediation message played an exaggerated role in our comparison of the messages, which may not make a big difference in a real context.

\emph{Fourth}, our experimental design assumes that there are always 4 messages, and that the mediator should intervene next. It does not include the messages after the intervention, or the important choice on when to intervene (c.f. \cite{branting2023computational}), 

While these choices were made to enable the assessment of LLMs in mediation, they also somewhat limit the general applicability of the results.
Future work should focus on evaluating such tools in real-world contexts, and involve expert mediators, in order to achieve a higher ``construct validity,'' i.e., be more closely aligned with real-world outcomes (c.f. \cite{kapoor2024promises}). 

\section{Conclusion \& Future Work}

In this study, we demonstrated that large language models possess significant potential in mediating disputes, performing on par or even surpassing our human annotators in selecting appropriate intervention types and crafting effective intervention messages. These findings suggest that LLMs could play a pivotal role in enhancing online dispute resolution platforms by providing scalable and cost-effective mediation services.

Our research contributes to the growing body of knowledge on AI applications in law and dispute resolution, highlighting the capabilities of LLMs in understanding complex human interactions and responding with empathy and neutrality. This advancement could significantly improve access to justice, particularly in cases where traditional mediation is inaccessible due to cost or availability constraints.

Future work should incorporate multi-modal data to better simulate real-world mediation scenarios, and conduct pilot studies within actual ODR systems to assess practical effectiveness. By continuing to refine these technologies, we move closer to a future where AI not only supports but enhances the human capacity for dispute resolution, contributing to a more accessible and efficient justice system.


\bibliography{ios-book-article}

\end{document}